\documentclass[a4paper,11pt]{article}
\usepackage{pos}

\title{Snowmass’21 Discussions on Future Accelerator HEP Facilities}

\author[a]{Stephen Gourlay}
\author[b]{Tor Raubenheimer}
\author*[c]{Vladimir Shiltsev}
\affiliation[a]{Lawrence Berkeley National Laboratory, Berkeley, CA 94720, USA}
\affiliation[b]{SLAC National Laboratory, Menlo Park, CA 94025, USA}
\affiliation[c]{Fermi National Accelerator Laboratory,  PO Box 500, Batavia, IL 60510, USA}

\emailAdd{shiltsev@fnal.gov}

\abstract{
The US particle physics community planning exercise (a.k.a. “Snowmass”) is organized every 7 to 9 years to provide a forum for discussions among the entire particle physics community to develop a scientific vision for the future of particle physics in the U.S. and its international partners. The Snowmass'21 Accelerator Frontier activities include discussions on high-energy hadron and lepton colliders, high-intensity beams for neutrino research and for “Physics Beyond Colliders”, accelerator technologies, science, education and outreach as well as the progress of core accelerator technologies, including RF, magnets, targets and sources. Here we summarize the Snowmass'21 discussions on future HEP accelerator facilities.}

\FullConference{%
  41st International Conference on High Energy physics - ICHEP2022\\
  6-13 July, 2022\\
  Bologna, Italy

}

\begin{document}
\maketitle

For almost a century, high-energy particle accelerators have played a key role in shaping modern nuclear and particle physics. 
Many large accelerators are currently under construction -- High Luminosity LHC upgrade, PIP-II, NICA, XFELs, EIC, ESS, FAIR, etc -- and will become operational in due time. We have also witnessed great progress toward frontier facilities for neutrino and rare processes physics research,  Higgs factories (linear or circular), and multi-TeV  $pp, \mu\mu$ and $e^+e^-$ colliders -- all of which are in the focus of the Snowmass'21 discussions. 

\section{Snowmass Process}

Snowmass is a particle physics community study that takes place in the US every 7-9 years (the last one  was in 2013). The Snowmass'21 study (the name is historical, originally held in Snowmass, Colorado) is organized by the Divisions of Particles and Fields (DPF), Beam Physics (DPB), Nuclear Physics (DNP), Astrophysics (DAP) and Gravitation (DGRAV) of the American Physical Society. Snowmass'21 strives to define the most important questions for the field and to identify promising opportunities to address them, see  https://snowmass21.org/. It provides an opportunity for the entire particle physics community to come together to identify and document a scientific vision for the future of particle physics in the U.S. and its international partners. The P5, Particle Physics Project Prioritization Panel, chaired by Hitoshi Murayama (UC Berkeley), will take the scientific input from Snowmass'21 (final summaries to be available in September 2022)  and by the Spring of 2023 develop a strategic plan for U.S. particle physics that can be executed over a 10 year timescale in the context of a 20-year global vision for the field. 

Snowmass'21 activities are managed along the lines of ten "Frontiers": 
Energy Frontier (EF), Neutrino Physics Frontier (NF), Rare Processes and Precision Frontier (RPF), Cosmic Frontier (CF), Theory Frontier (TF),  Accelerator Frontier (AF),  Instrumentation Frontier (IF), Computational Frontier (CompF),  Underground Facilities (UF), and Community Engagement Frontier (CEF). The Snowmass Community Summer Study workshop will take place in Seattle on July 17-26 and is the culmination of the various workshops and Town Hall meetings that have taken place during 2020, 2021, and 2022 as part of Snowmass’21. More than three thousand scientists have taken part in the Snowmass'21 discussions and about 1400 people will participate in the Seattle workshop in person and remotely. In general, the international community was very well represented and many scientists from Europe and Asia have been either organizers of sessions and events, or conveners of topical groups, or submitted numerous Letters of Interest (short communications) or White Papers (extended input documents). 

The key questions for the AF include:  \emph{What is needed to advance the physics? 
What is currently available (state of the art) around the world? 
What new accelerator facilities could be available in the next decade (or next next decade)? 
What R\&D would enable these future opportunities? 
What are the time and cost scales of the R\&D and associated test facilities, as well as the time and cost scale of the facilities?} There are 9 AF topical groups led by internationally recognized researchers: AF1 "Beam Physics and Accelerator Education" - Mei Bai (SLAC), Zhirong Huang (SLAC), Steve Lund (MSU); AF2	"Accelerators for Neutrinos" -  John Galambos (ORNL), Bob Zwaska (FNAL), 	Gianluigi Arduini (CERN); AF3 "Accelerators for EW/Higgs" -  Angeles Faus-Golfe (IN2P3), Georg Hoffstaetter (Cornell), 	Qing Qin (ESRF), Frank Zimmermann (CERN); AF4 "Multi-TeV Colliders"  - Mark Palmer (BNL), Nadia Pastrone (INFN), Jingyu Tang (IHEP), Alexander Valishev (FNAL); AF5 "Accelerators for Physics Beyond Colliders and Rare Processes" - Mike Lamont (CERN), Richard Milner (MIT), Eric Prebys (UC Davis); AF6 "Advanced Accelerator Concepts" -  Ralph Assmann (DESY), Cameron Geddes (LBNL), Mark Hogan (SLAC), Pietro Musumeci (UCLA);  AF7 "Accelerator Technology - RF" - Emilio Nanni (SLAC), Sergey Belomestnykh (FNAL), Hans Weise (DESY); AF7 "Accelerator Technology - Magnets" -	Susana Izquierdo Bermudez (CERN),	Gianluca Sabbi (LBNL), Sasha Zlobin (FNAL); AF7 "Accelerator Technology - Targets/Sources" - Charlotte Barbier (ORNL), Frederique Pellemoine (FNAL), Yin-E Sun (ANL). 

\section{Snowmass'21 Accelerator Frontier Discussions: Facilities}

More than 300 Letters of Interest and 116 White Papers have been submitted to the Snowmass'21 AF topical groups.  There were more than 30 topical workshops, 8 cross-Frontier {\it Agoras} (5 on various types of colliders: $e+e- / \gamma \gamma$, linear/circular, $\mu \mu$, $pp$, advanced ones and three on experiments and accelerators for rare processes physics), and several special cross-Frontier groups were organized between  AF, EF, TF, IF, and NF such as {\it the eeCollider Forum, the Muon Collider Forum, the Implementation Task Force} (see below), the 2.4MW proton power upgrade design group at FNAL, etc. While the topical group and frontier conveners are still putting the final touches on their reports, here we would like to emphasize several important outcomes of the Snowmass process:

{\bf Facilities for Neutrino Frontier:} 
The most powerful accelerators for neutrino research to date are the rapid cycling synchrotron facilities; J-PARC in Japan that has reached 515 kW of the 30 GeV proton beam power, and the Fermilab Main Injector delivering up to 893 kW of 120 GeV protons on target. They support neutrino oscillations research programs at the SuperK experiment (295 km from J-PARC) and MINOS (810 km from Fermilab), correspondingly. 

The needs of neutrino physics call for the next generation, higher-power, megawatt and multi-MW-class superbeams facilities. Elements of the LBNF/DUNE Phase II, a world leading neutrino experiment, 
were discussed at Snowmass'21. There is a broad array of accelerator and detector technologies and expertise to design and construct a 2.4 MW beam power upgrade of the Fermilab accelerator complex for Neutrino Frontier studies \cite{PIPIII}, expand the volume of Liquid Argon detectors by 20 ktons, and build a new neutrino near-detector on the Fermilab site. 

{\bf Facilities for Rare Processes Frontier:} Several possibilities for Rare Processes Frontier (searches for axions, charged lepton flavor violation, dark matter) have been identified that call for broad use of existing and future facilities, such as the SLAC 4-8 GeV electron linac, Fermilab’s PIP-II proton linac beam \cite{PAR}, etc.

{\bf Facilities for the Energy Frontier:} 
Charged particle colliders -- arguably the most complex and advanced scientific instruments --  have been at the forefront of scientific discoveries in high-energy and nuclear physics since the 1960s \cite{RMP2020}. There are seven colliders in operation and the Large Hadron Collider now represents the "accelerator energy frontier" with its 6.8 TeV energy per beam, 2.1$\cdot 10 ^{34}$ cm$^{-2}$s$^{-1}$ luminosity and some one TWh of annual total site electric energy consumption. The Super-KEKB is an asymmetric $e^+e^-$ B-factory with 4 and 7 GeV beam energies, respectively. Since the startup in 2018, it has achieved the world record luminosity (for any collider type) of 4.6$\cdot 10^{34}$ cm$^{-2}$s$^{-1}$, and aspires to reach 80$\cdot 10 ^{34}$ cm$^{-2}$s$^{-1}$-- a whopping 40-times over its predecessor KEK-B (1999-2010). 

\begin{figure}[htbp]
\centering
\includegraphics[width=0.66\linewidth]{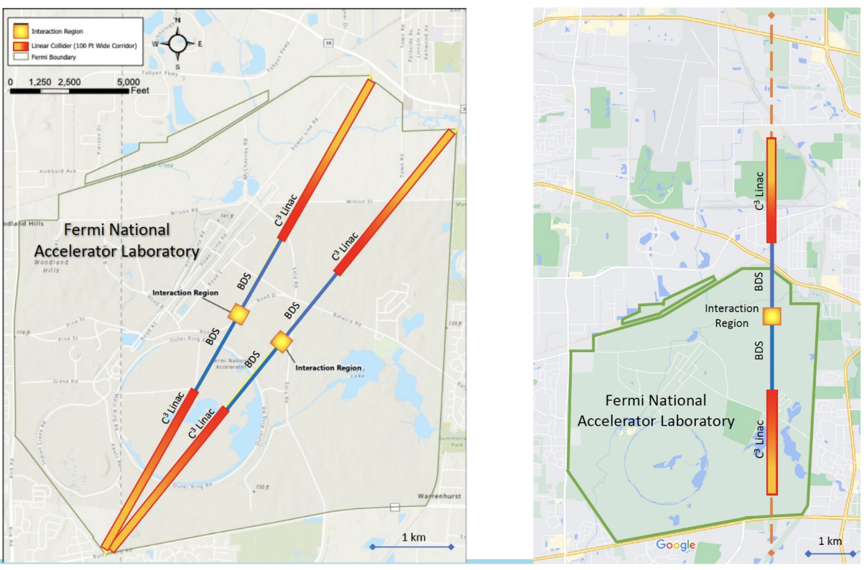}
\caption{Possible future linear $e+e-$ Higgs factory collider $C^3$ placement on the Fermilab map (from Ref.\cite{C3}). } 
\label{Fig1}
\end{figure}

The Energy Frontier community calls for an active program toward post-LHC colliders.  In particular, the world community has called for a Higgs/EW Factory as the next major accelerator project and this might be followed by a O(10 TeV/parton c.m.e.) collider. At present, there are as many as eight Higgs/EW factories under consideration --  $e^+e^-$ colliders such as the CEPC in China and FCCee at CERN, both roughly 100 km circumference, that require $O$(100 MW) RF systems to sustain high luminosity \cite{NatPhysCEPC}; or a 11 km long  CLIC (CERN) two-beam normal-conducting RF linear accelerator with average 72 MV/m gradient \cite{NatPhysCLIC}; or the 21 km long International Linear Collider (ILC) based on SRF 31.5 MV/m SRF linacs \cite{NatPhysILC}. Two recently proposed Higgs factories can be shorter than 7 km and potentially fit the Fermilab site - $C^3$ (employs 5.7 GHz 70 MV/m cool copper RF at 77 K) \cite{C3} and HELEN (1.3 GHz travelling wave SRF at 2K and 70 MV/m gradient) \cite{HELEN} - see Fig.\ref{Fig1}. 

In addition, there are also about two dozen energy frontier collider concepts that go beyond HL-LHC in their discovery potential. Among them are the 3 TeV CLIC option (100 MV/m accelerating gradient, 50 km long), a 10-14 TeV c.m.e. $\mu^+\mu^-$ collider (10-14 km circumference, 16 T magnets) \cite{NatPhys2020}, and two roughly 100 km circumference $pp$ colliders; SPPC in China (75-125 TeV c.m.e., based on 12-20 T IBS SC magnets) and FCChh at CERN (100 TeV, 16-17 T Nb$_3$Sn SC dipoles) \cite{NatPhysFCC}.

\begin{figure}[htbp]
\centering
\includegraphics[width=0.6\linewidth]{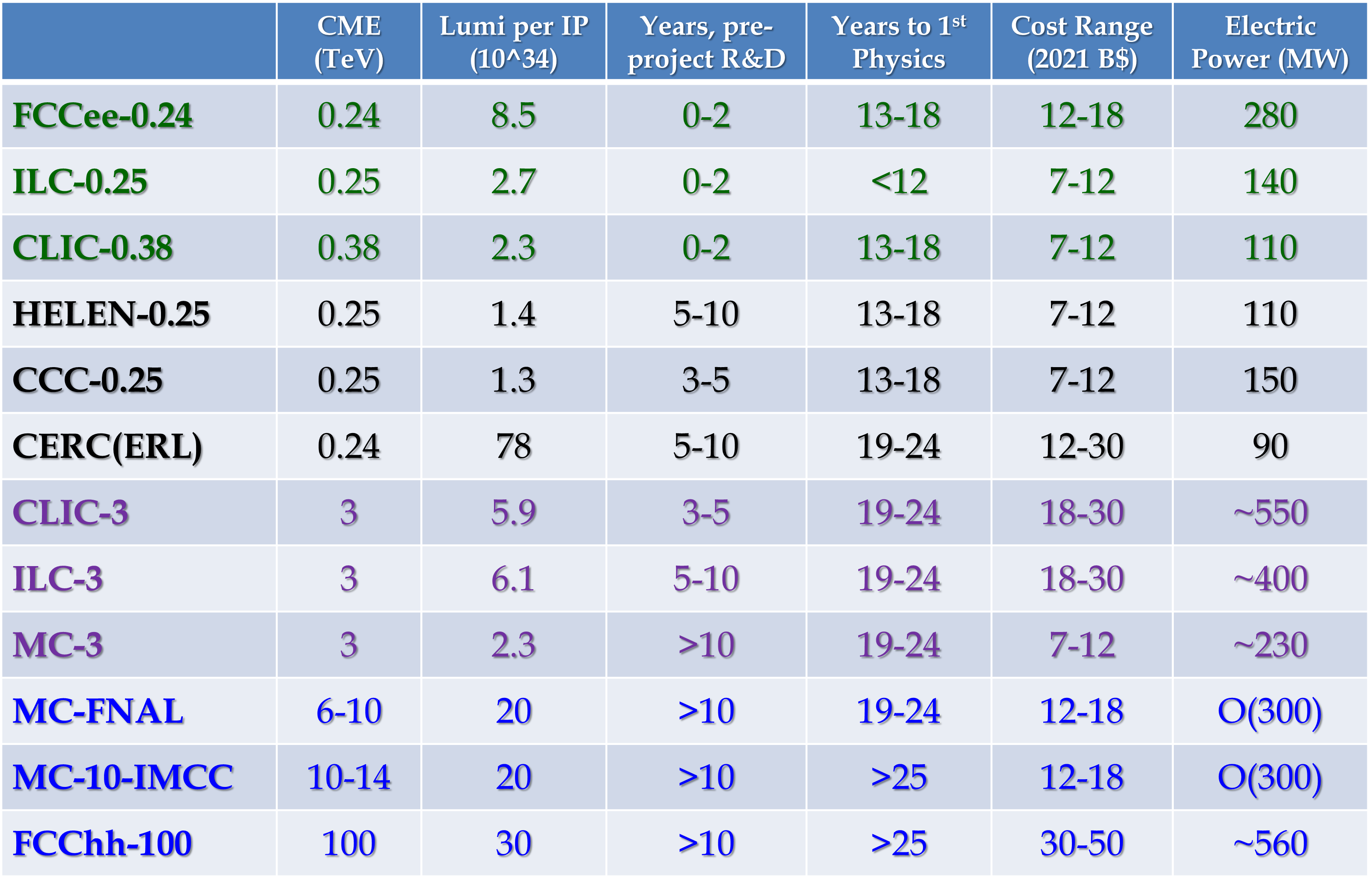}
\caption{Main parameters of the submitted Higgs factory proposals (FCCee, ILC, CLIC, $C^3$, HELEN, and CERC - ERL based collider in the FCCee tunnel) and multi-TeV colliders (CLIC, ILC, 3-, 10- and 14-TeV c.m.e. Muon Collider options, and FCChh). Years of the pre-project R\&D indicate required effort to get to sufficient technical readiness. Estimated years to first physics are for technically limited timeline starting at the time of the decision to proceed. The total project cost ranges are in 2021\$ (based on a parametric estimator and without escalation and contingency). The peak luminosity and power consumption values have not been reviewed by ITF and represent proponent inputs. (Adapted from the ITF draft report.)} 
\label{Fig2}
\end{figure}

{\bf Implementation Task Force:} 
The AF Implementation Task Force (ITF) was organized and charged with developing metrics and processes to facilitate comparisons between projects. The ITF was chaired by Thomas Roser (BNL) and was comprised of 11 additional world-renowned accelerator experts from Asia, Europe and the US including two  representatives of the \emph{Snowmass Young} (the Snowmass'21 organization of early career researchers), as well as three EF and TF liaisons. More than 30 collider concepts have been comparatively evaluated by the ITF using parametric estimators to compare physics reach (impact), beam parameters, size, complexity, power, environment concerns,  technical risk, technical readiness, validation and R\&D required, cost and schedule
 -- see Fig.\ref{Fig2}. 

{\bf Future Collider R\&D Program:} 
In the course of the AF discussions, clearly identified was the need of an integrated future collider R\&D program (a focused R\&D program in  the US DOE Office of HEP) to engage in the design and to coordinate the development of next generation collider projects such as: FCC-ee (circular collider), C3/HELEN/CLIC (linear Higgs factory colliders, the first two fitting the Fermilab site), multi-TeV Muon Collider, and FCC-hh, in order to enable an informed choice by the next Snowmass/P5 ca. 2030. The proposal of such a program will need to be approved by the P5 \cite{ColliderRandD}.

\section{Discussions on General Accelerator R\&D, Education, and Training}

The cost of large accelerators is set by the scale (energy, length, power) and technology. Typically, accelerator components (NC or/and SC magnets and RF systems) account for $50\pm10 \%$ of the total cost, while the civil construction takes $35\pm15 \%$, and power production, delivery and distribution technology adds the remaining $15\pm10 \%$ \cite{CostModel}. While the last two parts are mostly determined by industry, the magnet and RF technology is a linchpin of the progress of accelerators and would dominate the accelerator cost without progress from the R\&D programs. The U.S. has an active R\&D program in labs and universities aimed at general accelerator R\&D (GARD) and detectors that are critical in developing accelerator science and experimental technologies for future HEP accelerators (RF, magnets, beam physics, advanced concepts, targets and sources, beam physics, etc.) 

{\bf General Accelerator R\&D} 
 
Most general conclusions of the AF community discussions regarding aspirational goals for the GARD thrusts for the next decade include: a) in the area of targets and sources - development of efficient high intensity high brightness $e+$ sources and multi-MW proton targets for neutrino production (2.4 MW for PIP-III, 4-8 MW for a future muon collider); b) in the area of magnets for colliders and RCSs - design and testing of 16 T dipoles, 40T solenoids, and $O$(1000 T/s) fast cycling magnets (activities will be coordinated with the US Magnet Development Program); c) in the area of SC/NC RF - 70-120 MV/m $C^3$ and 70 MV/m TW SRF cavities and structures, exploration and testing of new materials with the potential of sustaining higher gradients with high $Q_0$, and development of efficient RF sources;  d) in advanced acceleration methods - the most needed are demonstration of collider quality beams, efficient drivers and staging, and development of self-consistent parameter sets of potential far-future colliders based on wakefield acceleration in plasma and structures (in close coordination with international programs such as the European Roadmap, EUPRAXIA, etc.); e) finally, in beam physics - the focus should be on experimental, computational and theoretical studies on acceleration and control of high intensity/high brightness beams,  high performance computer modeling and AI/ML approaches, and design integration and optimization, including the overall energy efficiency of future facilities.

{\bf Accelerator Workforce Education, Training, DEI and URM:} There is a recognized need to strengthen and expand education and training programs, enhance recruiting (especially international talent), promote the field (e.g., via colloquia at universities), and creating a national undergraduate level recruiting program structured to draw in women and underrepresented
minorities (URM), with corresponding efforts at all career stages to support, include and retain them in the field.

\vskip 0.1in

This manuscript has been supported by the Fermi Research Alliance, LLC under Contract No. DE-AC02-07CH11359 with the U.S. Department of Energy, Office of Science, Office of High Energy Physics and by the SLAC National Accelerator Laboratory under Contract No. DE-AC0376F00515 with the U.S. Department of Energy.

\end{document}